\documentclass[11pt, 3p,  review]{elsarticle}

\usepackage{natbib}

\usepackage{amsmath}   

\graphicspath{{./figures/}}
 
\DeclareGraphicsExtensions{.png}

\usepackage{booktabs}
\usepackage{textcomp}
\usepackage{tabularx}
\usepackage{arydshln}

\usepackage{lscape}

\usepackage{float}
\restylefloat{table}

\usepackage{color}
\usepackage{soul}

\begin{document}

\begin{frontmatter}
\title{Cooperation dynamics under pandemic risks and heterogeneous economic interdependence}

\author[ugr,uon] {Manuel Chica~\corref{cor}}
\ead{manuelchica@ugr.es}

\author[gc] {Juan M. Hern{\'a}ndez}
\ead{juan.hernandez@ulpgc.es}

\author[ist] {Francisco C. Santos}
\ead{franciscocsantos@tecnico.ulisboa.pt}

\address[ugr]{Andalusian Research Institute DaSCI ``Data Science and Computational Intelligence", \\ University of Granada, 18071 Granada, Spain}

\address[uon]{School of Electrical Engineering and Computing, \\ The University of Newcastle, Callaghan, NSW 2308, Australia}

\address[gc]{Department of Quantitative Methods in Economics and Management, \\
Universtiy Institute of Tourism and Sustainable Economic Development (TIDES), \\
University of Las Palmas de Gran Canaria, 35017 Las Palmas, Spain}

\address[ist]{INESC-ID \& Instituto Superior Técnico, Universidade de Lisboa, 2744-016 Porto Salvo, Portugal}

\cortext[cor]{Corresponding author}

\begin{abstract}

The spread of COVID-19 and ensuing containment measures have accentuated the profound interdependence among nations or regions. This has been particularly evident in tourism, one of the sectors most affected by uncoordinated mobility restrictions. The impact of this interdependence on the tendency to adopt less or more restrictive measures is hard to evaluate, more so if diversity in economic exposures to citizens' mobility are considered. Here, we address this problem by developing an analytical and computational game-theoretical model encompassing the conflicts arising from the need to control the economic effects of global risks, such as in the COVID-19 pandemic. The model includes the individual costs derived from severe restrictions imposed by governments, including the resulting economic interdependence among all the parties involved in the game. By using tourism-based data, the model is enriched with actual heterogeneous income losses, such that every player has a different economic cost when applying restrictions. We show that economic interdependence enhances cooperation because of the decline in the expected payoffs by free-riding parties (i.e., those neglecting the application of mobility restrictions). Furthermore, we show (analytically and through numerical simulations) that these cross-exposures can transform the nature of the cooperation dilemma each region or country faces, modifying the position of the fixed points and the size of the basins of attraction that characterize this class of games. Finally, our results suggest that heterogeneity among regions may be used to leverage the impact of intervention policies by ensuring an agreement among the most relevant initial set of cooperators.

\end{abstract}


\end{frontmatter}

\section{Introduction}
\label{sec:introduction}

The COVID-19 pandemic has caused the most severe global economic breakdown of the recent history, mainly due to the measures to control the virus spread (e.g., quarantines, stay-at-home and social distancing policies), which have produced a dramatic shutdown of the economic activity. In this context, the coordination among countries is an essential instrument to efficiently control pandemic and enhance economic recovery~\citep{Svoboda2020,Holtz2020}.\par

The pandemic control measures result in two types of negative economic effects on countries and regions. First, the direct effects, derived from internal mobility restrictions, which seriously injure many sectors in the country. They do not just include hospitality and entertainment, but also banks, stock market and education sector~\citep{Ozili2020}. Second, indirect effects, induced by trade and financial linkages among countries, which produce economic contagion of the consequences of restrictions in a country to a foreign linked country~\citep{Ozili2020,Fernandes2020}. An example of an indirect effect is the travel restrictions in tourist dependent countries, which produce a serious economic loss to destination countries~\citep{Pham2021}. Other pandemic consequences are found in international supply chains. Thus, export-oriented countries are influenced by demand shortfall of the importing countries and, at the same time, supply disruption from some countries such as China produces input shortages and inflationary pressure in import-countries~\citep{Pahl2021}. In addition, the contact pattern in some regions are influenced by the social distancing policies followed in others~\citep{Holtz2020}. All this phenomena stem from the economic interdependence (EI) among countries.  \par

Furthermore, the economic impact of the pandemic control measures is not homogeneous for all the countries and regions. Those service and export-oriented countries, such as those depending on tourism, entertainment and transport, have been more affected than others~\citep{world2020global}. For example, Panama, which depends mainly on tourism and transport services, suffered a serious GDP decrease in 2020 (19\%), and other many tourism-dependent small-islands states, such as Maldivas, Fiji and Bahamas, experimented similar economic consequences. In Europe, the COVID-19 pandemic economic loss in Spain, around 10\% decrease of the GDP, duplicated the economic loss in Germany. These evidences show that the economic consequences of the implementation of control measures are markedly heterogeneous.\par

The coordination problem of pandemic control measures can be represented through a collective-risk dilemma (CRD)~\citep{Milinski2008,Santos2011}. This is a multiplayer public good game (PGG) where every player can contribute with some amount to avoid a certain risk of failure. Normally, it is not necessary that all players contribute to achieve the common goal and there is some space for \textit{free-riders} who benefit from the others' contribution. Traditionally, cooperation to mitigate climate change by cutting carbon emissions has been one of the most important applications of CRDs~\citep{tavoni2011inequality,milinski2011cooperative,AbouChakra2012,Pacheco2014,Vasconcelos2014,Gois2019,domingos2020timing}. More recently, CRDs have been proposed for the coordination of restrictions and reopen policies in the current context of the COVID-19 pandemic~\citep{Chica2021}.  \par

The aim of our study is to analyze the conditions for achieving coordination among countries to control pandemic situations, taking into account the specific economic consequences of these measures. Specifically, we propose a new CRD model where EIs among the players are considered by altering the expected profits by regions given the restrictions applied by cooperating regions. The proposed CRD model also incorporates the heterogeneity found for regions and countries with respect to their income loss when cooperating by restricting the economic activity. Heterogeneity is omnipresent in reality but only few studies included heterogeneous features in social dilemmas~\citep{Li2021HeteroInteractions}.

We fit a log-normal distribution for feeding the model with heterogeneous economic loss values by analyzing real data from tourism contribution to GDP of the European Union NUTS2 regions. The experiments comprise the evaluation of the final cooperation levels of the population with and without EI and validate them under the presence of heterogeneity. Our results show that the existence of EI among countries can favor cooperation for all the tested conditions. To understand the reason behind these observations, we extract the stable and unstable points of the new dilemma with and without heterogeneity. To this end, we propose a new way of calculating the internal roots from the agent-based simulation results. \par

Finally, we take advantage of the reality coming from the income loss heterogeneity of the model to analyze those initial conditions facilitating cooperation. Three scenarios are evaluated. We first set the initial cooperators at random but we also condition those initial cooperators by a positive and negative bias by they income losses of the regions. The reader will see how significant differences in the final cooperation levels are achieved when choosing the best initial conditions alternative. These results will help to engineer more effective governmental policies to increase cooperation.

\section{\label{sec:model}Model}   

We present a CRD model to represent the cooperation game among regions or countries (we call them countries from now on) when adopting measures to control pandemic spread having into account their negative economic consequences. We resort to evolutionary game theory and stochastic population dynamics \citep{hofbauer1998evolutionary,sigmund2010calculus,nowak2004emergence,traulsen2006stochastic, nowak2006evolutionary,Perc10,Perc17} when required, combined with agent-based computer simulations \citep{Macal05,Adami16}. The inspiration for this model was found in previous CRD models for cooperative actions against climate change~\citep{Santos2011,Vasconcelos2014} and pandemic spread~\citep{Chica2021}.

\subsection{\label{sec:model_hom} Game definition including economic interdependence (EI)}

The model includes a finite number $Z$ of countries or regions. Every player $i$ can choose two strategies $s_i(t)$ at every time step $t$: cooperation ($C$), which means adopting measures to control the pandemic spread and suffering income loss; and defection ($D$), which means not adopting any measure, continuing with the normal economic activity and eventually free-riding the correct public health conditions derived from others' cooperation.   
 
Players interact in groups of size $N$, representing international or regional agreements, alliances or work-groups. By assumption, these groups are randomly formed. Every group faces a risk $r \in [0,1]$ for the health care system to collapse and the resulting economic breakdown, if the epidemic is not controlled enough inside the group. This happens when the number of cooperators in the group does not achieve a minimum $M \leq N$. When the number of cooperators is equal or higher than $M$ in a group of size $N$ countries, the pandemic is under control and the economic breakdown is not produced. 

When a country $i$ cooperates, economic activity critically stops and an income loss $c_i$ occur in the region ought to these restrictions. This loss is not necessarily constant throughout the population, but every country has its loss value, which depends on the specific economic structure of the country. In addition, the economic breakdown in the cooperator $i$ is spread to the rest of countries due to the international economic linkages among them. Thus, defector countries are not free of economic negative spillovers or contagion from other countries. Instead, they are influenced by the number of cooperators in the total population. The more the cooperator countries in the total population, the higher the negative economic consequences in the defector country. 

The conditions above are represented in the expected payoff $\Pi_D^i$ ($\Pi_C^i$) of a defector (cooperator) country $i$. They are: 

\begin{eqnarray} \Pi_D^i(j) &=& \left(1-c_i\frac{k}{Z}\right)[\Theta(j-M) + (1-r) \left(1- \Theta (j-M)\right)], \label{payoff-hom1}\\
\Pi_C^i(j) &=& \Pi_D^i(j) - \left(1-\frac{k}{Z}\right) c_i,
\label{payoff-hom2}
\end{eqnarray}

\noindent where $j$ is the number of cooperators in the group and $k$ is the number of cooperators in the total population. The variable $M=mN$, where $N$ is the group size and $m$ is the minimum fraction of cooperators to avoid collapse. The Heaviside step function $\Theta(x)$ is equal to $0$ whenever $x<0$
and equal to $1$ otherwise. The initial endowment or maximum payoff obtained in absence of any pandemic is normalized to $1$.

The first term of Equation \ref{payoff-hom1} shows that when the number of cooperators in the group is above the threshold $M$, the expected payoff of a defector is not maximum, but lowers an amount due to the economic interdependence with cooperator countries. Specifically, the loss of the defector's income is a fraction $\frac{k}{Z}$ of its own maximum loss $c_i$. This term $c_i\frac{k}{Z}$ is added to the cooperator's payoff (Equation \ref{payoff-hom2}) since we assume that the single cost of cooperation $c_i$ already includes the economic contagion derived from international linkages. The total loss in the cooperators' income is $c_i$, as expected. When the necessary number of cooperators in a group is not achieved ($j<M$), both cooperators and defectors have a risk $r$ of global economic collapse and null economic activity.


\subsection{\label{sec:ev_dynamics}Individual update of the game strategies} 

After playing a game round $t$, players can update their strategies according to the received payoffs. Here, we have considered the Fermi function as the evolutionary update rule~\citep{szabo1998evolutionary,Traulsen2006}. The Fermi rule is a stochastic pairwise comparison rule in which strategies that do well, are more likely to be imitated, and spread throughout the population. {In detail, at each time-step, a player $i$ with a payoff $\Pi_{i}$ is randomly selected from the population for strategy revision. Player $i$ will then randomly select another player $j$ from the population as a potential role model; $i$ will imitate the strategy of $j$ with a probability $p$ that increases with their payoff difference --- ($\Pi_{j} - \Pi_{i}$) --- and can be written as in \citep{traulsen2006stochastic}}:

\begin{equation}
p = \frac{1}{1 + e^{- \beta(\Pi_{j} - \Pi_{i})}}.
\label{eq:fermi}
\end{equation} 

The free parameter $\beta$ is the intensity of selection, encoding the chance of mistakes during the imitation process. This means that a player $i$ can copy another player's strategy $j$ despite having a lower payoff. We set $\beta=1$ in all the experiments of the study. Additionally, the players of the game can randomly explore other strategies, adopting a strategy at random with probability $\mu$. This mutation (or exploration) probability ($\mu$) equals $\frac{1}{Z}$ in all experiments. This update rule can be used in both synchronous and asynchronous paradigm. We have confirmed via numerical simulations that our conclusions remain valid for both cases.

\subsection{\label{sec:transition_model}Stochastic population dynamics} 


Let us start by assuming that the income loss $c_i$ is homogeneous for all players. Then we have a unique income loss $c$ for all countries and the payoff derived from any strategy depends on the general parameters and the number of cooperators $j$ in the group. Let us also assume that all players are equally likely to interact, a configuration known as a well-mixed population~\citep{sigmund2010calculus}. In other words, we have a random selection of partners to form groups. In this limit, we can analytically compute the expected payoff (or fitness) of a cooperator/defector for a given number of cooperators $k$ in the population by using an hyper-geometric sampling~\citep{hauert2006synergy,Pacheco2009}: 

\begin{eqnarray*} f_C(k) &=& \left( \begin{array}{c} Z-1 \\N-1  \end{array} \right)^{-1} \sum_{j=0}^{N-1} \left( \begin{array}{c} k-1 \\j  \end{array} \right)\left( \begin{array}{c} Z-k \\N-j-1  \end{array} \right)  \Pi_C(j+1), \\
f_D(k) &=& \left( \begin{array}{c} Z-1 \\N-1  \end{array} \right)^{-1} \sum_{j=0}^{N-1} \left( \begin{array}{c} k \\j  \end{array} \right)\left( \begin{array}{c} Z-k-1 \\N-j-1  \end{array} \right)  \Pi_D(j),
\end{eqnarray*}

\noindent where we have removed the super-index referring to player $i$ in the payoff functions. 

The update rule described in section \ref{sec:ev_dynamics} defines a Markov process where, at every time step, the probability to increase the number of cooperators $k$ in the population is \citep{traulsen2006stochastic},
\begin{equation}
T^{+}(k) = \frac{Z-k}{Z}\left[(1-\mu)\frac{k}{Z}\frac{1}{1 + e^{- \beta(f_C(k)-f_D(k))}}+\mu \right],
\label{eq:T+}
\end{equation} 
and the probability to decrease the number of cooperators is
\begin{equation}
T^{-}(k) = \frac{k}{Z}\left[(1-\mu)\frac{Z-k}{Z}\frac{1}{1 + e^{- \beta(f_D(k)-f_C(k))}}+\mu \right].
\label{eq:T-}
\end{equation} 

Several tools can be used to analyze the evolutionary dynamics emerging from this 
ergodic Markov chain. First, the gradient of selection \citep{Pacheco2009,Santos2011}, $G(k)=T^{+}(k)-T^{-}(k)$, indicates the direction of change for every cooperation level. Second, as the process includes probabilistic mutation ($\mu$), the population does not fixate in any stationary state. Thus, instead of computing the probability of fixation in each absorbing state, we can make use of the stationary distribution of this Markov chain to analyze the asymptotic state of the population, and assess the pervasiveness in time of a each fraction of cooperators. To this end, we build the transition matrix $M=\left(p_{ij}\right)_{(Z+1)\times(Z+1)}$, where $p_{k,k\pm 1}=T^{\pm}(k)$, $p_{k,k}=1-p_{k,k+1}-p_{k,k-1}$ and 0 for the rest. This is a tridiagonal matrix and the stationary distribution is the normalized eigenvector $(\bar{p}_k)_{k=0..Z}$ corresponding to eigenvalue 1 of the transpose of matrix $M$. Third, the expected final number of cooperators is calculated from the stationary distribution, as $n_C=\sum_{k=0}^Z\bar{p}_kk$. We can also use the hypergeometric distributions above to calculate the probability $a_G(k)$ of having groups with $M$ cooperators or more for every cooperation level $k$ \citep{Vasconcelos13,Vasconcelos2014}. This is given by $\eta_G = \sum_{k=0}^Z\bar{p}_ka_G(k)$ indicating the average group achievement of the game in the stationary solution. \par

\subsection{\label{sec:computer_simulations}Agent-based computer simulations} 

While convenient for the case with homogeneous income losses ($c$), the analytical mean-field approach described above is no longer valid in the case of heterogeneous exposures or losses. The inclusion of regions or countries' diversity in exposure introduces a higher level of complexity may, nonetheless, be conveniently described through  Monte-Carlo (MC) agent-based simulations~\citep{Macal05,Adami16}, performed in computer clusters and resorting to parallel computing architectures. Evolution proceeds in discrete steps involving imitation and mutation, in line with the stochastic dynamics described above.

We fix the size of the groups $N$ to $10$ and the minimum threshold $m$ to $0.7$ (i.e., $M=7$) for the whole set of experiments shown in the study. Experiments with group sizes $N=\{5, 10, 25\}$ and $m=\{0.5, 0.7\}$ were carried out without noticing significant changes. For the sake of numerical tractability, the size of the population in the analytical results is $Z=200$. For the agent-based simulations, the size of the population is $Z=2.0\times10^3$ and 
we run the model for $30$ independent MC realizations and a maximum number of $10^3$ synchronous time-steps, where all the realizations reach a stationary stable state and deviation from the MC realizations is low. Finally, all the simulation results were obtained by averaging the last 25\% of the simulation time-steps in the independent MC runs.

\section{\label{sec:results}Results and discussion} 

In this section we analyze the results of three different experiments by considering the effects of the economic interdependence (EI) in the dilemma. First, we present the analytical study of the game when having homogeneous costs $c$ in Section~\ref{sec:increase_coop_homo}. We later apply agent-based simulations to the dilemma with costs heterogeneity (Section~\ref{sec:increase_coop_hetero}) and study the roots of the dilemma in Section~\ref{sec:barriers_points}. Finally, Section~\ref{sec:fixing_initial_conditions} provides guidance on how to engineer interventions for increasing cooperation by taking into account heterogeneity.


\subsection{\label{sec:increase_coop_homo}Analytical study of the cooperation increase when considering EI} 

\begin{figure}[htb!]
\centering
\includegraphics[width=0.7\textwidth]{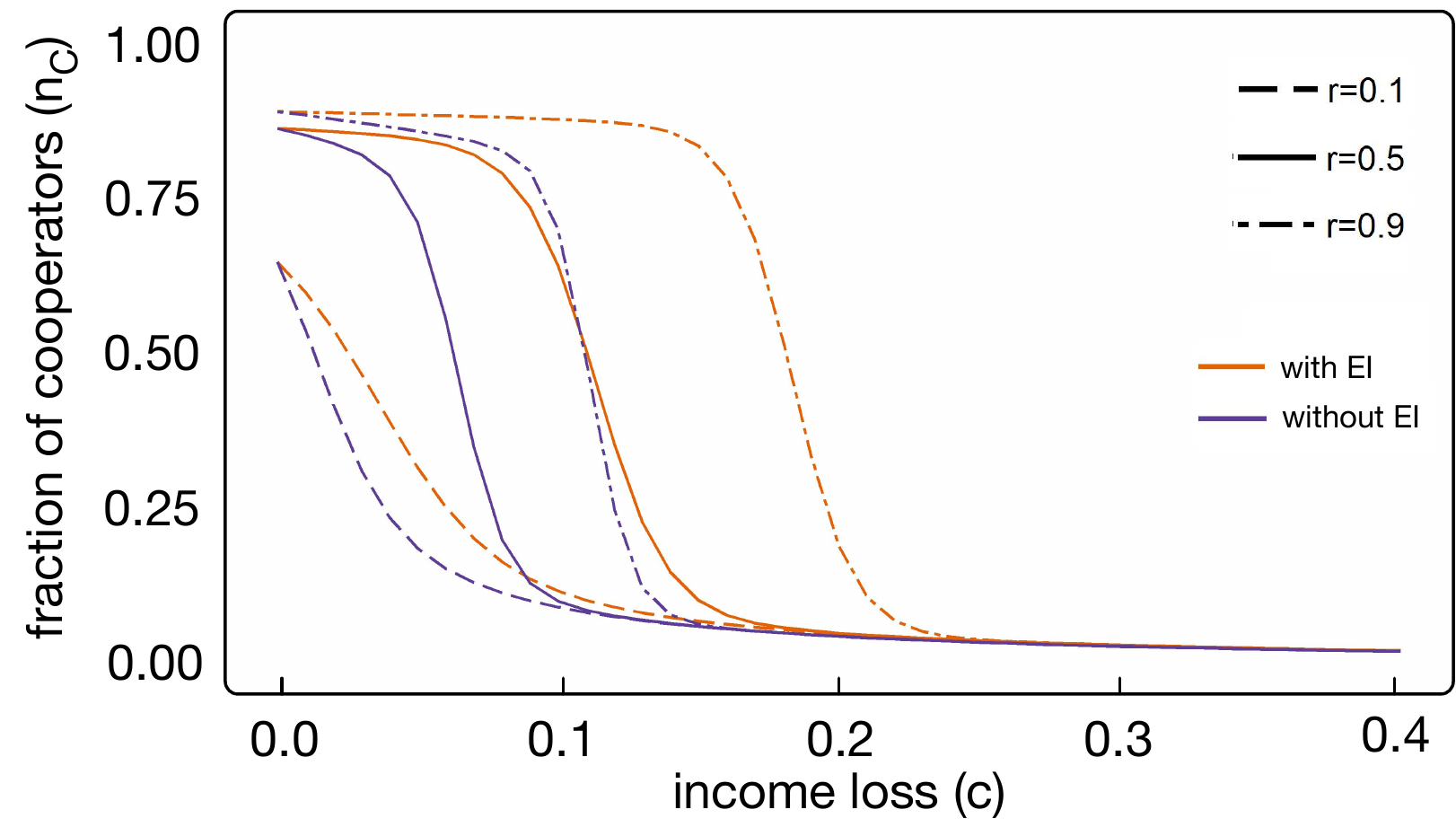}
\caption{Final fraction of cooperators for different values of homogeneous income loss $c$ for the models with economic interdependence (EI) and without EI. We assume three risk levels. This figure shows that EI increases the levels of cooperation and group success for wide intervals of income losses and risk values. The parameter values are: $Z=200$, $N=10$, $M=7$, $\beta=1$, $\mu=1/Z$.} 
\label{fig:homo_costs}
\end{figure}

First, we analyze the effect of the income loss on the final stationary state of the game. Figure \ref{fig:homo_costs} shows the average final percentage of cooperators as a function of the homogeneous income loss $c$ for three risk levels. To analyze the effect of economic interdependence (EI), we represent the trajectories with two models: one adopting the payoff equations \ref{payoff-hom1} and \ref{payoff-hom2} (model with EI) and other one using the same equations but removing the term $c_i\frac{k}{Z}$ (model without EI).\par

As it can be observed, the income loss negatively influences on the final number of cooperators. In general, its effect is not strong for income loss values near zero, but up to a certain threshold, the expected number of cooperators dramatically decreases until achieving almost null values. The lower the risk associated with the game, the lower the levels of income loss where null cooperation is achieved.

Moreover, Figure \ref{fig:homo_costs} shows the positive effect of EI on the cooperation in the three risk levels shown. Therefore, the existence of economic linkages among countries in the pandemic context favors cooperation in the dilemma. The gap in the number of cooperators between the two models is larger for larger risk values. This result is apparently counter-intuitive, since including new economic costs to the players may in general deter the disposition to cooperate. However, this is not the case in this context, since the income loss due to EI reduces the overall economic benefit at stake for all players and therefore shortens the benefit margin of being defector.

\begin{figure}[htb!]
\centering
\includegraphics[width=0.7\textwidth]{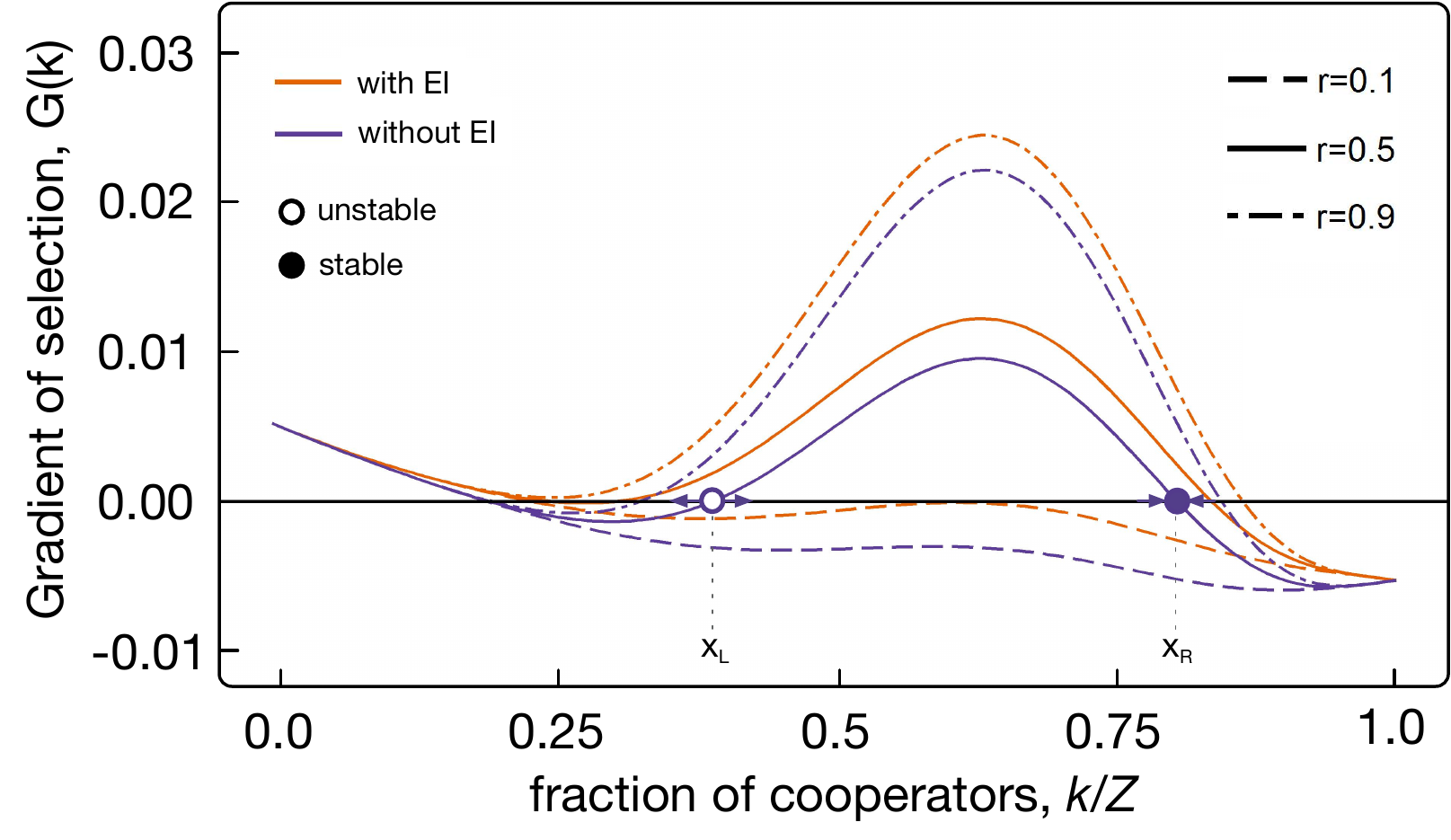}
\caption{Gradient of selection for the models with and without EI. 
According to the sign of the Gradient of selection $G(k)$, cooperators (defectors) are likely to increase in the population whenever $G(k)>0$ ($G(k)<0$). We assume the same three risk levels as in Figure ~\ref{fig:homo_costs}. EI is able to transform the overall dynamics, by reducing the coordination requirements ($x_L$) required to reach a cooperative basin of attraction. Similarly, EI also increases the stable fraction of cooperation ($x_R$) reached whenever $x_L$ is overcome. Parameter values: $Z=200$, $N=10$, $M=7$, $\beta=1$, $\mu=1/Z$.}
\label{fig:gradient_homo}
\end{figure}

A deeper insight of the analytical solution of the model with and without EI is presented in Figure ~\ref{fig:gradient_homo}, which shows the gradient of selection $G(k)$ for the same three risk levels in Figure \ref{fig:homo_costs}, with and without EI. If the gradient of selection, $G(k)$, is positive (negative), the number of cooperators is likely to increase (decrease) whenever the population has $k$ cooperators. The roots of $G(k)$ offer the finite population analogues of fixed points in infinite populations \citep{Pacheco2009,Santos2011}. Here, we can identify configurations with two internal roots typical of this class of dilemmas: one unstable root on the left-hand side ($x_{L}$) and one stable root ($x_{R}$) for higher fractions of cooperators, associated with two well-defined basins of attraction. It also suggests that a critical mass of cooperators ($x_{L}$) needs to be surpassed such that the system naturally self-organizes into a co-existence of cooperators and defectors.

Interestingly, the gradients in Figure ~\ref{fig:gradient_homo} show that stable equilibria ($x_{R}$) occur for higher levels of cooperators when EI is included. At the same time, the coordination point ($x_{L}$) tends to move towards lower values of the fraction of cooperators $k/Z$ when EI is in place. This suggests that EI reduces the requirements to sway to the cooperative basin of attraction where cooperators and defectors may co-exist. Naturally, the position of both ($x_{L}$) and $x_{R}$ depends on the value of risk ($r$). The higher the perception of the risk, the lower the unstable equilibrium $x_{L}$ and the higher the stable equilibrium $x_{R}$, being more likely to overcome the dilemma.  The position of these roots --- but also the amplitude of $G(k)$ --- determine the final stationary distribution and the expected prevalence of cooperators. Finally, the exploration (or mutation) probability ($\mu$) drive the system to the center of the cooperation axis by making $G(0)>0$, further favouring the shift towards the right-hand side of the simplex.

\subsection{\label{sec:increase_coop_hetero}Simulation-based results for heterogeneous income losses based on real data} 

\begin{figure}[htb!]
\centering
\includegraphics[width=0.7\textwidth]{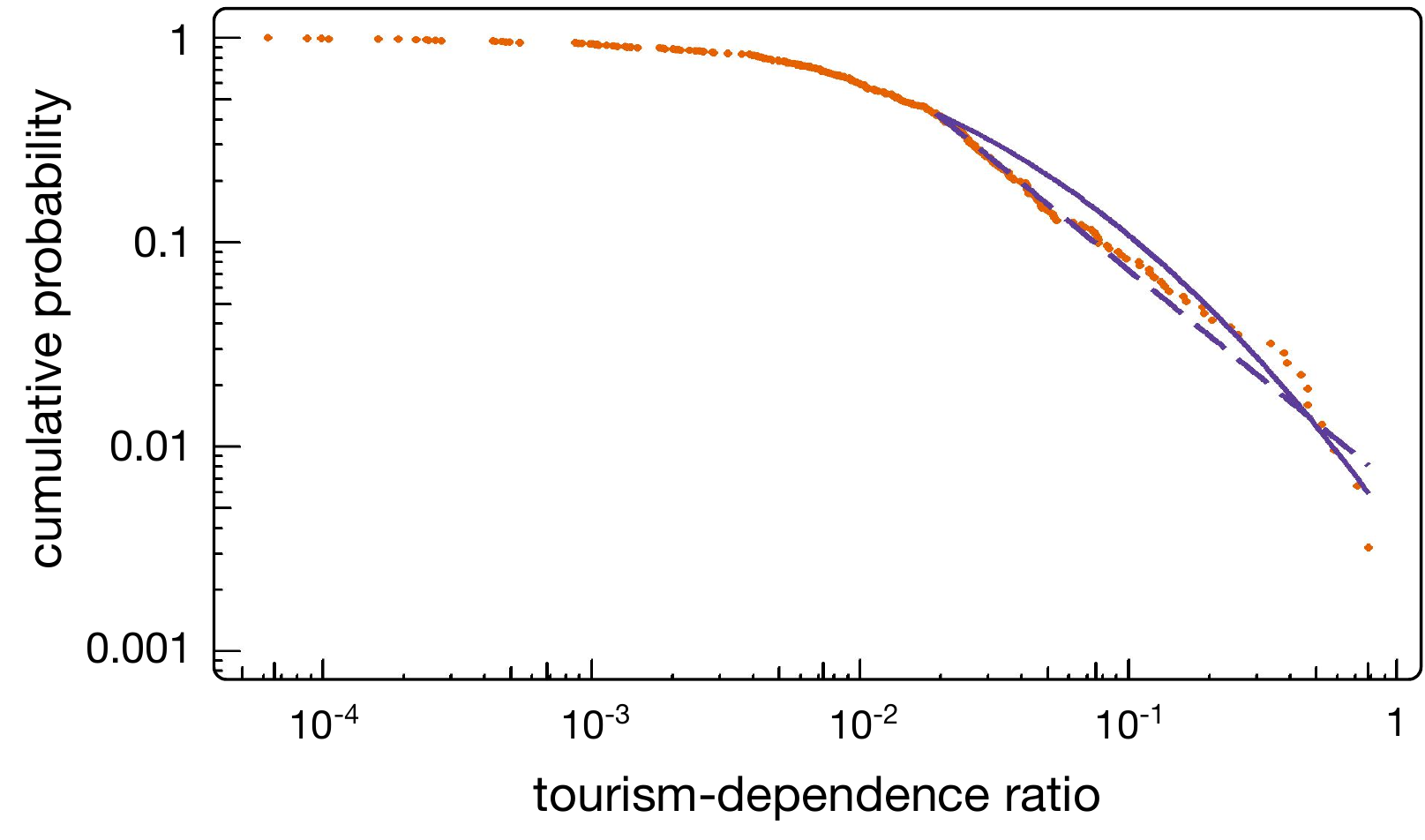}
\caption{Cumulative distribution of the income loss for the 312 regions of the EU NUTS2 (circles). It shows that regions' typically portray a significant heterogeneity in the dependence from tourism activity. Blue solid line represent the fitted log-normal distribution and the dashed line the fitted Pareto distribution (see main text for details). }
\label{fig:distribution}
\end{figure}

As commented in the introduction, the income loss due to the implementation of pandemic control measures is not homogeneously distributed among countries. For example, tourism was one of the most affected sectors by the COVID-19 pandemic and tourism GDP contribution of the European regions is one example of the heterogeneity of the players involved, as discussed in~\citep{Chica2021}. The number of regions in the EU NUTS2 classification~\citep{EurostatNights} is 312 and we use real data from this classification to compute the exposure of the regions to a lockdown and tourism activity halt. The considered indicator for this contribution are the nights spend at tourist accommodation establishments per inhabitant. Although the averaged contribution of the regions is $0.04$, we observe a clear heterogeneity in the distribution. Figure~\ref{fig:distribution} shows the 312 data points and two fitted distribution: a log-normal and a power-law. 

The log-normal distribution with fitted parameters $\mu_{ln} = - 4.39$ and $\sigma_{ln} = 1.63$ is the one with the best fitting. Therefore, we use this distribution to feed the costs or income loss
$c_i$ for the players. Because of the heterogeneity in the costs, it is not possible to fully assess the overall dynamics through the mean-field analysis of the previous section. Thus, in the following, we resort to agent-based simulations to estimate the evolutionary dynamics of the model~\citep{Adami16,Chica2021}. We have tested both log-normal and power law distributions of potential income losses without significant changes and therefore, the same conclusions apply for both distributions. 

\begin{figure}[htb!]
\centering
\includegraphics[width=0.85\textwidth]{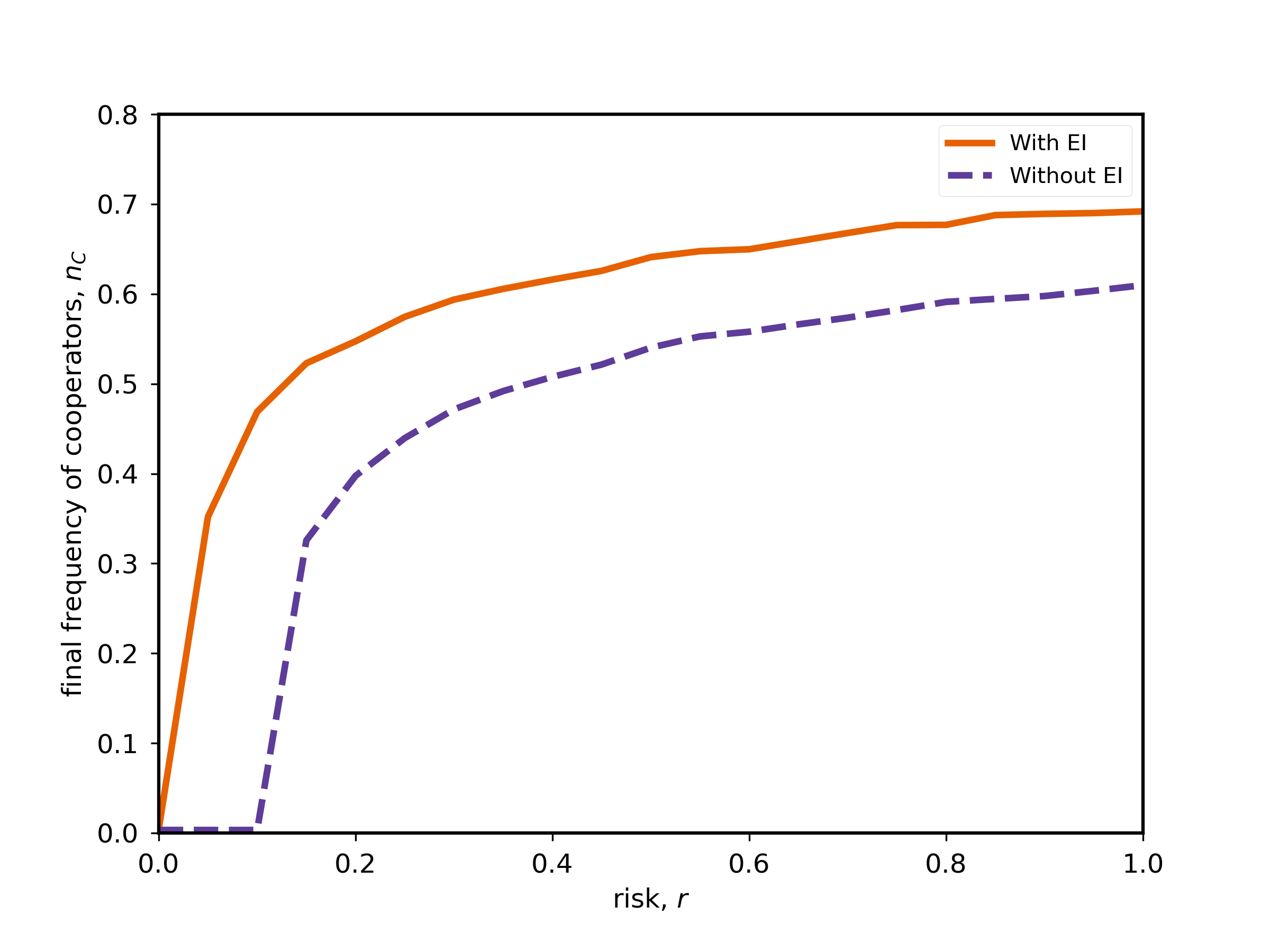}
\caption{Final fraction of cooperators $n_C$ for different risk levels $r$ comparing the dilemma with and without economic interdependence when having heterogeneous income loss from a fitted log-normal distribution. Economic interdependence (EI) leads to high levels of cooperation, even when regions portray a large diversity in economic exposure to confinement measures. Results obtained via numerical simulations with $Z=2\times10^3$, $N=10$, $M=7$,  $\beta=1$, and $\mu=1/Z$.}
\label{fig:cooperation_with_without_EI}
\end{figure}

Figure~\ref{fig:cooperation_with_without_EI} shows the averaged final cooperation frequency for the game with and without EI using the log-normal distribution of income loss. To obtain each point of both curves at every risk value, we average the simulations results from a set of sufficient discrete values for the initial frequency of cooperators of the system, $n^0_C$. 

We can see in this figure how the increase in cooperation is clear for all the risk levels after averaging all the possible initial conditions of cooperation. The inclusion of EI boosts cooperation even for very low risk values and the positive gap in cooperation remains practically equal for the whole range of risk values. Therefore, we see that the main behavior of the system is the same when injecting heterogeneity through the real distribution of income loss $c_i$. The results are in line with the homogeneous setting of the model. The global increase in final cooperation when incorporating EI in the dilemma is robust, independently from the heterogeneity of the income loss.

\subsection{\label{sec:barriers_points}Study on the internal fixed points of the game}

\begin{figure}[htb!]
\centering
\includegraphics[width=0.9\textwidth]{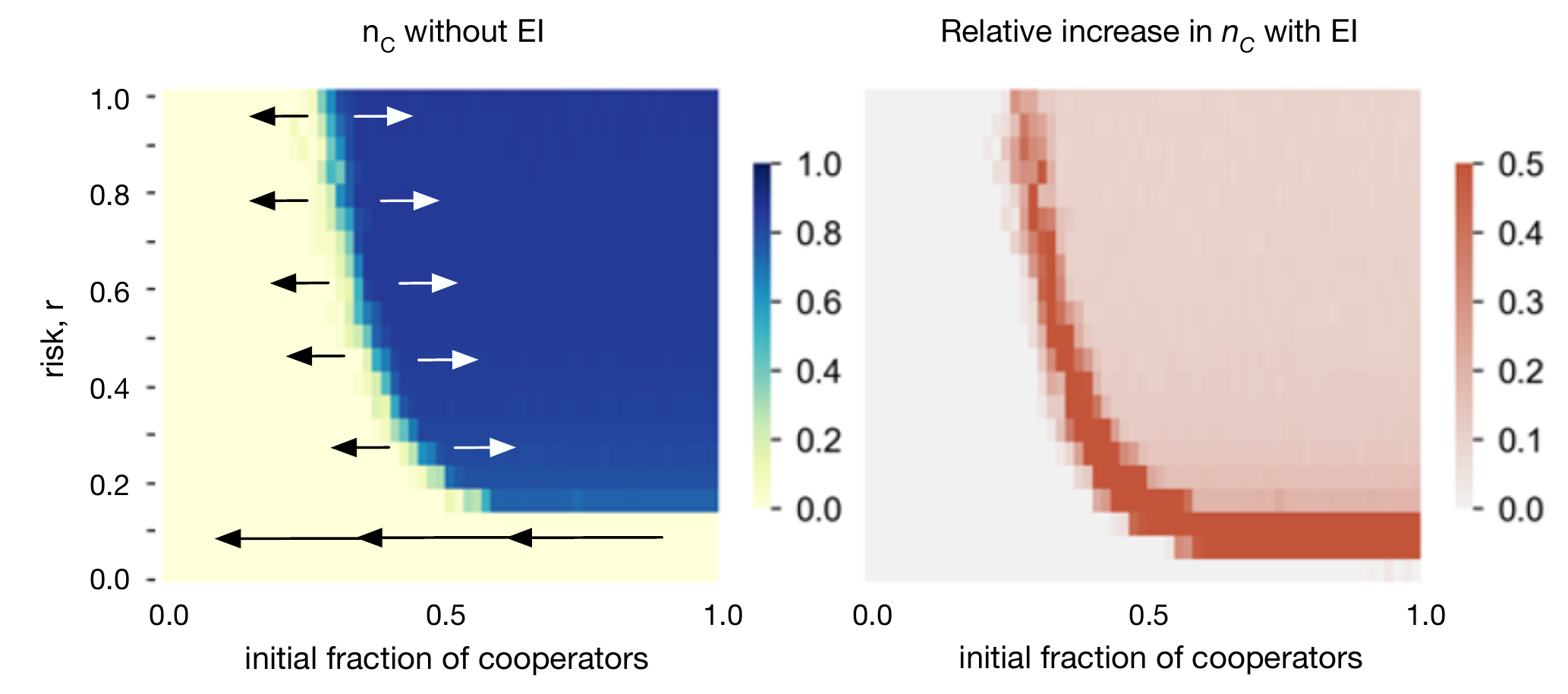}
\caption{Heatmaps comparing the sensitivity analysis on $r$ and the initial fraction of cooperators when having a log-normal heterogeneous distribution. Left plot shows the absolute $n_C$ when not considering EI while right plots shows the relative increase of the dilemma when considering EI with respect to the CRD without EI (left plot). Results obtained numerically, with the same parameters of Figure \ref{fig:cooperation_with_without_EI}.}
\label{fig:heatmaps_lognormal}
\end{figure}

The adoption of a different exposure for each agent makes it hard to assess the overall dynamics from an analytical perspective. Particularly, the inference of the effective gradient of selection and associated equilibrium from large-scale computer simulations is not always trivial (see, e.g.,\citep{pinheiro2016linking}). In this section we aim at explaining why there is an increase in cooperation when simulating the model with heterogeneous costs and estimate the roots of the dynamics as in the analytical model. Therefore, we first propose a methodology to estimate, from the sensitivity analysis on risk $r$ and initial cooperators $n^0_C$, the stable and unstable equilibrium points. Figure~\ref{fig:heatmaps_lognormal} shows this sensitivity analysis on $r$ and $n^0_C$ where each cell is the averaged final frequency of cooperators $n_C$ from the MC realizations of the agent-based model using the log-normal distribution of income loss $c_i$. The plot on the left represents the absolute $n_C$ values of the game without EI while the plot on the right shows the relative increase in $n_C$ of the game including EI with respect to the traditional dilemma.

From the heatmaps in Figure \ref{fig:heatmaps_lognormal}, we see a clear increase in cooperation for the majority of $r$ values. We may also take profit from the knowledge obtained in Figure \ref{fig:gradient_homo} regarding the existence of two internal fixed points to, analogously, try to find these roots in the heterogeneous case. As before, the heatmaps portray two basins of attraction: one in which cooperation cannot be sustained (yellowish areas) and other where cooperators and defectors co-exist (blue areas). In order to estimate the stable and unstable points from the heatmaps, we follow the next method:

\begin{itemize}

    \item Unstable fixed points ($\kappa_L$): for every value of $r$, we calculate the $n^0_C$ value where the increase in final cooperation $n_C$ is maximum. We discard small steps because of numerical simulation variations such as those when $r$ is low (e.g., below $0.1$ when no EI, in Figure~\ref{fig:heatmaps_lognormal}). 
    
    \item Stable fixed points ($\kappa_R$): for every value of $r$, we select the lower $n^0_C$ value where the majority of the players of the population are cooperators. This is done by defining a minimum threshold for $n_C$ to consider the population as cooperator. We set this threshold to $0.85$ for extracting the stable points of our experiments.
    
\end{itemize}

\begin{figure}[htb!]
\centering
\includegraphics[width=0.8\textwidth]{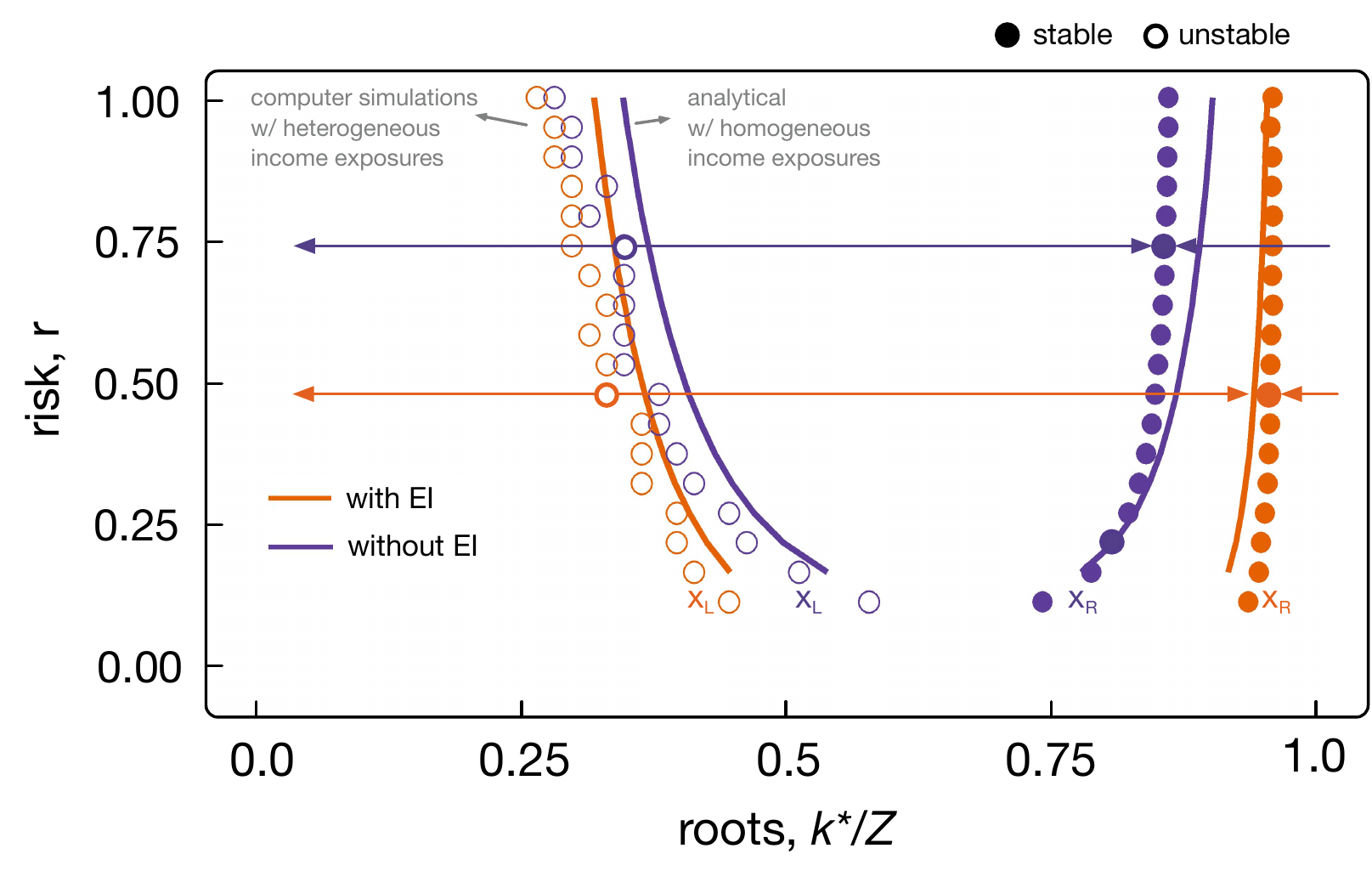}
\caption{Analogues of unstable (empty circles) and stable (full circles) roots obtained from computer simulations with log-normal distribution of heterogeneous costs. Purple and orange lines show the position of the stable and unstable roots obtained analytically for homogeneous income exposures. Purple lines and points represent the dilemma without EI while orange colour is for the dilemma with EI. Other parameters: $Z=2\times10^3$, $N=10$, $M=7$,  $\beta=1$, and $\mu=1/Z$.}
\label{fig:barrier_points}
\end{figure}

Figure~\ref{fig:barrier_points} shows fixed points for the population dynamics in both analytical and simulation-based approaches, with and without EI. Empty circles represent unstable fixed points ($\kappa_L$), and full circles represent stable fixed points ($\kappa_R$), coming from the simulation-based outputs of the heterogeneous setting, following the above-mentioned method. Lines are calculated from the analytical approach when using homogeneous costs (i.e., $c_i = 0.04, \forall i \in Z$).

If we compare both analytical and simulation-based points, we see results are in line even when having different heterogeneity settings. The incorporation of EI when there are pandemic risks that can affect the payoffs of the players by others' flow restrictions clearly shifts the fixed points. Stable points are shifted to the right and unstable points are shifted to the left when including this EI effects. The increase in cooperation is given by this extension of the internal points in the dilemma. We also show that the method to obtain the internal points by exploiting the simulation results is robust and the points are equivalent to the ones obtained by the analytical approach, obtained the same conclusions. Therefore, heterogeneity and the use of simulation techniques are not generating significant differences and results are solid.

\subsection{\label{sec:fixing_initial_conditions} Using heterogeneity to boost cooperation by fixing the initial cooperation conditions} 





Given the reality is heterogeneous and the fact we were able to incorporate this heterogeneity in the model by using simulation-based techniques, our aim is to exploit this information to glimpse ways of boosting cooperation. These insights can serve as a kick-off for employing policies by institutions. In this experiment we have fixed the initial conditions of the fraction of cooperators $n^0_C$ based on the heterogeneous values of cost or income loss of the countries or players $c_i$. 

Specifically, we have defined three main scenarios. In the first one, the initial cooperators are selected at random from the members of the population and there is not any initial bias. For the second one, we start fixing the initial cooperators from high to low values $c_i$. And finally, the third scenario considers the initial cooperators from low to high values of $c_i$. Thus, the second scenario has a positive bias of initial cooperation with respect to cooperation costs $c_i$ while the third scenario has a negative bias.

\begin{figure}[htb!]
\centering
\includegraphics[width=0.85\textwidth]{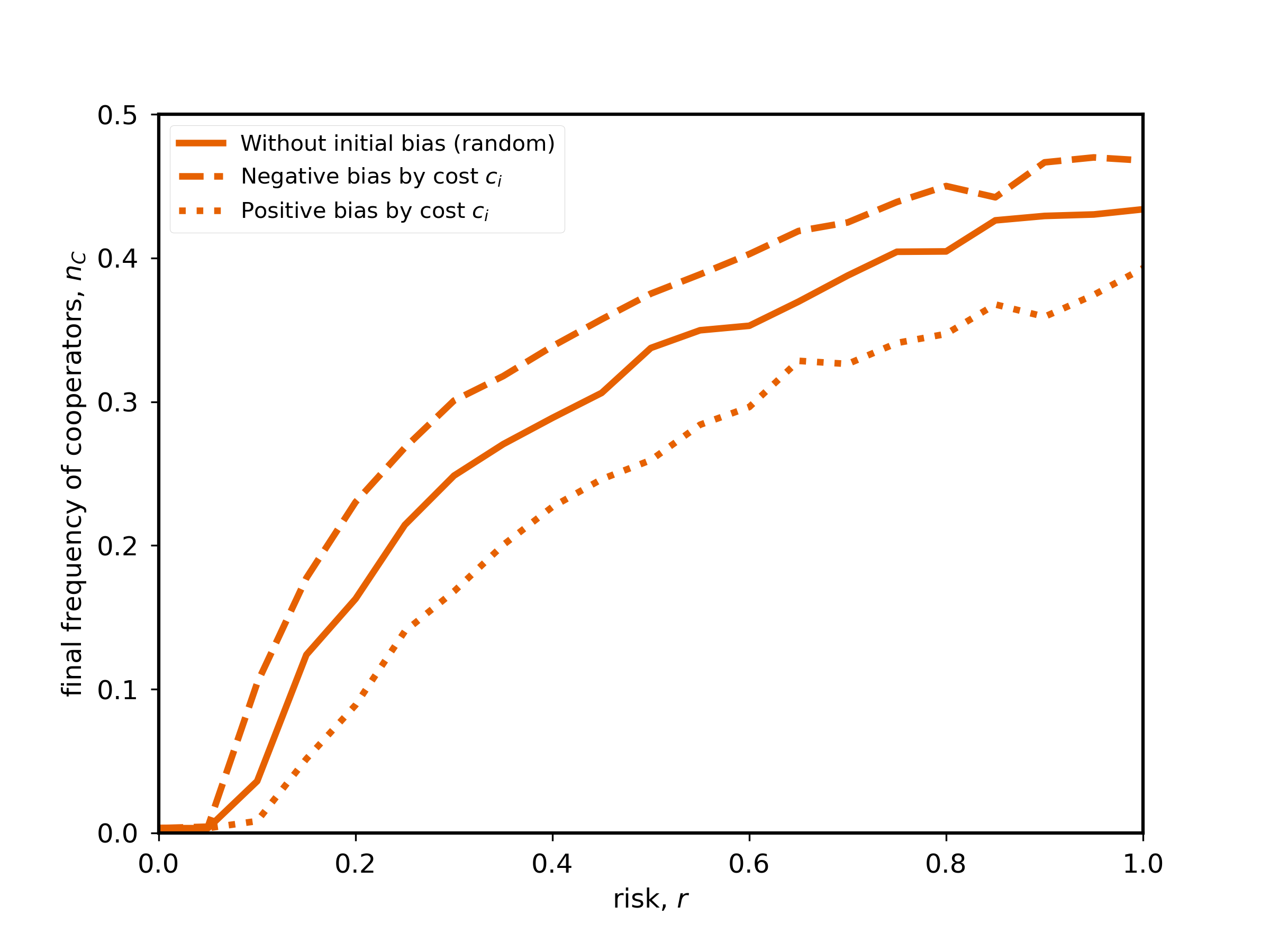}
\caption{Comparison on different initial conditions $n^0_C$ for achieving the best final frequency of cooperation $n_C$ in the population. Initial coordination of players with the lowest income loss can drive the entire population to higher levels of cooperation. On the opposite, if the initial cooperators face a significant income loss, cooperation tends to decrease compared with the random case. This result shows that heterogeneity leads to different liabilities, depending on the level of exposure. Moreover, it suggests that one may profit from heterogeneity to trigger prosociality at the global level. Parameters: $Z=2\times10^3$, $N=10$, M=7, $\beta=1$, and $\mu=1/Z$.}
\label{fig:initial_conditions}
\end{figure}

Figure~\ref{fig:initial_conditions} shows the $n_C$ comparison for a sensitivity analysis on risk $r$ using the three scenarios for the dilemma with EI using heterogeneous income loss (i.e., running the agent-based simulations). In order to get the lines of the plot we averaged a sufficient set of values for initial frequency of cooperators $n^0_C$ by following each of the three scenarios. The set of values is from $n^0_C=0$ to $n^0_C=0.5$ as policies should be focus on a reduced number of players in the population and it is not suitable for the comparison to restrict initial cooperators to a high number.
 
The results of the plot are clear. When the system starts with those players having the lowest income loss as cooperators, the final cooperation of the population increases significantly. On the opposite, when we fix the initial cooperators to those having the highest income loss, the final cooperation even decreases with respect to the random initial cooperators setting. Final cooperation increase of negative bias with respect to positive bias is between $10\%$ and $50\%$ depending on the risk values. Just when risk values are below $0.1$, all the scenarios have equal results as the dilemma does not facilitate any final cooperation.

\section{\label{sec:conclusions}Conclusions}


COVID-19 pandemic and other global risks have changed how regions or countries interact concerning global failures such as economic knockdowns. In these cases, there are economic interdependences (EIs) or spillover effects among players when facing public goods games. For instance, economic or mobility restrictions set by some players can affect the expected payoff by defecting ones or \textit{free-riders}. To cope with this new reality, we proposed a collective risk dilemma (CRD) that includes EI effects among players. Real data from the tourism contribution of the EU regions is employed to enrich the model by setting a genuine heterogeneous distribution of cooperation costs for the players. 

We show that EI robustly increases cooperation for both homogeneous and heterogeneous cases. EI is able to modify the (finite population analogues of) internal fixed points when compared with the dynamics in the absence of EI. We depart from a classic CRD characterized by defector dominance and coexistence among cooperators and defectors, each outcome associated with two well-defined basins of attraction. In the absence of any additional community enforcement mechanism, EI drops the minimum number of cooperators required to reach the cooperative basin of attraction, and increase the prevalence of cooperators in coexistence point. We have computed these fixed points analytically for scenarios with homogeneous costs and through agent-based simulations in the case of heterogeneous costs. To this end, in the latter, we proposed a new method to infer these finite population analogues of stable and unstable fixed points in infinite populations from the simulation's outputs. 

Finally, we have discussed how biased initial conditions based on the level of exposure may alter the final expected outcome. Results showed that the entire population benefits from having cooperators within the sub-group of players with lower income loss. This result suggests that one may profit from heterogeneity in designing effective interventions or governance policies to trigger prosociality at the global level, a result of particular importance if we consider that individual strategies may also depend on the perceived risk by each party \citep{amaral2021epidemiological}. Interventions should focus on those players showing a lower exposure to the economic risk. 

Future work can assess if rewarding and sanctioning activities can be applied~\citep{Vasconcelos13, Chen15, Gois2019, Couto2020,vasconcelos2020coalition, Hu2020rewarding} to a specific target sub-population and the features of this subset of individuals to target, or how positive, and negative incentives can be optimally distributed among groups and actors. Moreover, reactions to the COVID-19 pandemic have shown a wide range of (often polarized) responses. Recent results have shown how uncertainty may influence how each individual perceives the dilemma \citep{dannenberg2015provision,tavoni2011inequality,domingos2021modeling,barfuss2020caring}, potentially leading to polarized reactions \citep{domingos2020timing}, a development yet to be studied in the context of cooperation dynamics related to managing economic losses under pandemic conditions. Finally, how leaders act and influence others by their example and reputation can affect the whole population outcome~\citep{Wang2017leadership}, and this phenomenon can be studied for this dilemma. All these open questions remain critical in the current quest of understanding and promoting human cooperation, given the difficulty in assessing the advantages and disadvantages of each possible type of intervention policies.

 \section*{Acknowledgments} 

M.C. is supported by the Spanish Ministry of Science, Andalusian Government, and ERDF under grants SIMARK (P18-TP-4475), RYC-2016-19800, and Jose Castillejo program (CAS19/00090). J.M.H. is supported by the University of Las Palmas de Gran Canaria under grant COVID-19 04. F.C.S. acknowledges the support from FCT-Portugal (grants UIDB/50021/2020, PTDC/MAT-APL/6804/2020, and PTDC/CCI-INF/7366/2020).


\bibliographystyle{elsarticle-harv}

\end{document}